\documentclass{aa}
\usepackage{psfig}
\usepackage{graphics}
\newcommand{\rxj}{RX\,J0513}
\begin{document}


\thesaurus{06     
           (02.01.2;  
            08.02.1;  
            08.02.4;  
            08.09.2 RX J0513.9--6951;  
            08.14.2;  
            13.25.5)  
}
\title{A limit-cycle model for the binary supersoft X-ray source
RX\,J0513.9-6951}

\author{K. Reinsch \inst{1} \and A. van Teeseling\inst{1} \and 
	A. R. King\inst{2} \and K. Beuermann\inst{1} }

\institute{Universit\"ats-Sternwarte, Geismarlandstr. 11, 37083 G\"ottingen, 
	Germany \\ e-mail reinsch@uni-sw.gwdg.de
	\and 
	Astronomy Group, University of Leicester, Leicester LE1 7RH, U.K.
}

\date{Received November 30, 1999; accepted December 29, 1999}

\maketitle

\begin{abstract}

We present new results of our X-ray monitoring of the transient binary
supersoft X-ray source RX\,J0513.9--6951 in the LMC and of our
re-analysis of optical light curves obtained during the MACHO
project. We have covered a complete X-ray outburst cycle with the
ROSAT HRI detector. From the amplitude and timescale of the soft X-ray
variability, tight limits are derived for the temporal behaviour of
the white-dwarf radius and the effective temperature of its
envelope. A limit-cycle model is proposed to explain the observed
optical and X-ray variability, the characteristic timescales of the
durations of the X-ray on and off states, and those of the transitions
between both states. Our observations confirm that the radius changes
of the white-dwarf envelope occur on the Kelvin-Helmholtz
timescale. The duration of the X-ray on and off states is compatible
with the viscous timescales of the inner and outer accretion disk,
respectively.

\end{abstract}

\section{Introduction}

Luminous supersoft X-ray sources (SSS) have been established as a new and
distinct class of objects which are observationally distinguished by their
very soft X-ray spectra with temperatures on the order of 30\,eV and
luminosities of $10^{36} - 10^{38}$\,erg\,s$^{-1}$ (for recent reviews see
Kahabka \& van den Heuvel 1997; van Teeseling 1998). 
Several SSS have been identified as accreting close binaries with orbital
periods of $\sim$ 1 day or less. The most popular interpretation of these
systems involves a white dwarf which accretes matter via Roche-lobe
overflow and an accretion disk at a rate of $\sim 1-4 \times 10^{-7}
M_{\sun}$/yr, sufficient to permit stable quasi-steady nuclear
shell-burning in the surface layers of the white dwarf, either because of
thermal timescale mass transfer from a more massive (slightly evolved) main
sequence companion (van den Heuvel et al. 1992) or because of wind-driven
mass transfer from a low-mass irradiated companion (van Teeseling \& King
1998). 

\section{The transient binary SSS RX\,J0513.9--6951}

The luminous transient soft X-ray source RX\,J0513.9--6951 (henceforth
\rxj{}) discovered in the ROSAT all-sky survey (Schaeidt et al. 1993) has
been optically identified as a high mass-transfer accreting binary system
in the LMC (Cowley et al. 1993; Pakull et al. 1993) with an orbital period
of 0.76 days (Crampton et al.  1996). Optical monitoring has revealed that
\rxj\ undergoes recurrent low states at quasi-regular intervals, in which
the optical brightness drops by $\sim$ 1 magnitude (Reinsch et al. 1996;
Southwell et al. 1996). The optical low-states are accompanied by a turn-on
of the system in the soft X-ray range (Reinsch et al. 1996; Schaeidt 1996). 

The optical low states last for $\sim$ 40 days and repeat about every
140--180 days. Such short time scales cannot be explained by the
limit-cycle behaviour sketched by van den Heuvel et al. (1992) or by
recurrent burning models (Fujimoto 1982). Within the framework of a
shell-burning white dwarf an alternative explanation has been suggested by
Pakull et al. (1993): The rather sudden changes in the soft X-ray flux are
the result of the direct response of the white dwarf to slight changes in
the mass transfer rate. On the horizontal shell-burning branch, a small
increase of the accretion rate may significantly affect the effective
radius of the white dwarf envelope (Kato 1985). An increase of the
photospheric radius by e.g. a factor of 4 implies that the effective
temperature drops by a factor of 2. Given the extreme sensitivity of the
ROSAT PSPC and HRI count rates on temperature, this in turn implies that
the source may become undetectable although the bolometric luminosity
remains roughly the same (e.g. Heise et al. 1994). In this model of an
expanding and contracting envelope the sudden drop of the optical flux, the
colour variation, and the temporarily increased soft X-ray flux can be
quantitatively described by variations in the effective temperature of the
hot central star and variations in the irradiation of the accretion disk
(Reinsch et al. 1996). 

This model is supported by an independent estimate of the neutral hydrogen
column density obtained with recent HST UV spectroscopy which constrains
the X-ray luminosity during the on-state of \rxj\ to $(2.5 - 9) \times
10^{37}$\,erg\,s$^{-1}$, i.e. somewhat below the Eddington limit, and
confirms that the radius of the soft X-ray source is consistent with the
radius of a non-expanded white dwarf (G\"ansicke et al. 1998). 

\begin{figure} 
\resizebox{\hsize}{!}{\includegraphics{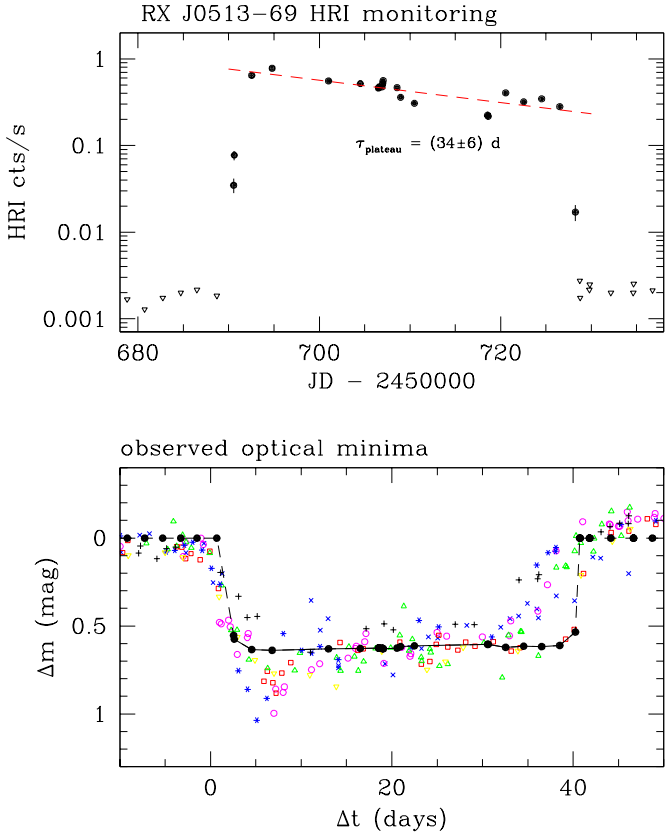}}
\caption[]{\label{hri_opt} 
X-ray and optical variability of the transient binary \rxj{}. Upper
panel: ROSAT HRI detected (filled circles) and upper limit count rates
(triangles) of \rxj\ during a complete outburst cycle. Lower panel: light
curves of different optical minima determined from CCD images taken for the
MACHO project. Filled circles predicted optical light curves calculated 
from our X-ray data (see text). }
\end{figure}

\section{X-ray and optical monitoring}

In order to obtain tight limits for the temporal development of the
radius and the effective temperature of the photosphere, we have
monitored \rxj\ with the ROSAT HRI detector at intervals of about two
days covering one complete X-ray outburst cycle (Fig. \ref{hri_opt},
upper part). The source remained undetectable during the X-ray off
state and showed a sudden increase of the soft X-ray flux by a factor
of $>100$ at the end of August 1997, reaching maximum flux after $\sim$ 5
days. The X-ray outburst lasted $\sim$ 40 days and ended with a steep
flux decline by again a factor of $>100$ to non-detectability within 2
days. The slow flux variation during the X-ray outburst can be
approximated by an exponential decline with a time constant of $(34\pm
6)$\,days.

\begin{figure}
 \psfig{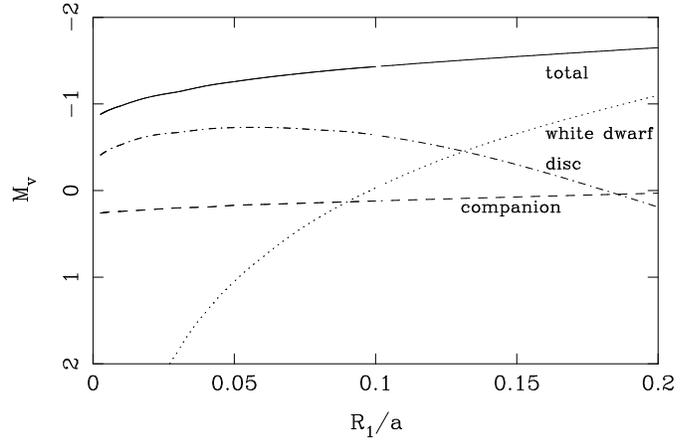}
\caption[]{\label{mvradius} Absolute $V$ magnitude as a function of
the white dwarf photospheric radius $R_1$ for the disk, the companion
star, the white dwarf, and their sum.}
\end{figure}

Detailed light curves of several optical low states are available from
regular monitoring of \rxj\ as a serendipitous source on CCD images taken
for the MACHO project (Southwell et al. 1996). In Fig. \ref{hri_opt} (lower
part) the light curves of different observed optical minima are shifted in
time such that the onset of the optical decline occurs at $\Delta t = 0$.
MACHO data obtained quasi-simultaneously with our HRI monitoring indicate
that the steep onset of the X-ray outburst occurred within 1--2 days after
the beginning of the optical decline (W. Sutherland, private
communication). 

The shape of the optical low states repeats fairly well with an
uncertainty of some 5 days in the duration of the low state. Within
this uncertainty, its length coincides with
the duration of the X-ray outburst. Both, the X-ray and the optical
light curves show a similar small flux gradient before the final fast
transition to the X-ray off/ optical high state occurs.

\section{Light curve modelling}

To test whether the model of a contracting and expanding white dwarf can
quantitatively explain both the X-ray light curve and the dips in the
optical light curve we have calculated a predicted optical light curve from
our X-ray data. First we used LTE white dwarf model atmosphere spectra (van
Teeseling et al.  1994) to determine the photospheric radius as a function
of the HRI count rate, where we assumed a distance of 50\,kpc, a bolometric
luminosity of $10^{38}$\,erg\,s$^{-1}$, and an absorption column of $n_{\rm
H} = 6\times 10^{20}$\,cm$^{-2}$ (G\"ansicke et al. 1998). Then we used the
binary light curve code {\sc binary++} (van Teeseling et al. 1998) to
calculate the orbital average optical magnitude as a function of the
photospheric radius $R_1$ of the white dwarf. Since this code self-consistently
calculates the amount of irradiation from an extended white dwarf on the
accretion disk and companion, including all possible shielding effects,
this calculation is more accurate than the semi-analytic approach we used
in Reinsch et al. (1996) and allows us to investigate how the results
depend on the various parameters. We assume a mass ratio of
$M_2/M_1 = 2$, an orbital separation $a = 3.8\times 10^{11}$\,cm as
appropriate for a quasi-main-sequence donor star and an orbital period $P =
0.76$\,days, an orbital inclination of $10\degr$, 
a disk filling 80\% of the average Roche-lobe radius, a uniform
irradiation reprocessing efficiency of $\eta=0.5$, a secondary
temperature of $9000$\,K, and an accretion rate of $3.4\times
10^{-7}\,M_{\sun}$.

Figure~\ref{mvradius} shows the resulting total absolute $V$ magnitude
as a function of $R_1$, and individual magnitudes of the disk, the
companion star and the white dwarf. For $R_1/a > 0.13$ or \mbox{$R_1
\ga 5\times 10^{10}$\,cm}, the expanded white dwarf is the dominant
optical light source. With increasing $R_1$ the disk first becomes
brighter because of more effective irradiation, but becomes fainter
again for $R_1 \ga 2\times 10^{10}$\,cm because an increasing part of
the inner disk disappears inside the white dwarf envelope.

\begin{figure}  \resizebox{\hsize}{!}{\includegraphics{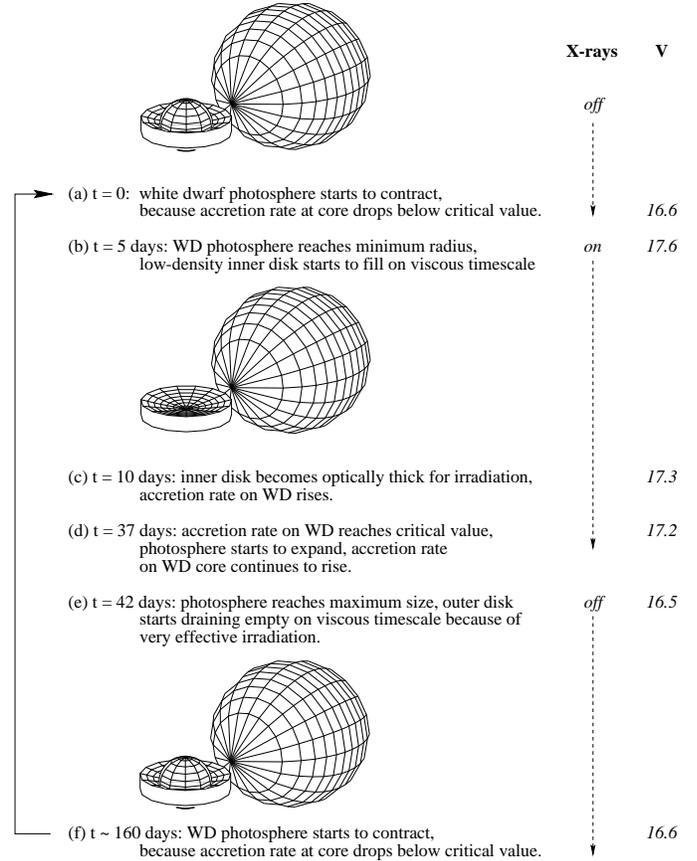}}
\caption[]{\label{scenario} Limit-cycle model for \rxj\ (see text).}
\end{figure} 

In Figure~\ref{hri_opt}, we have plotted the predicted optical light
curve over the combined MACHO light curve. For data points with only
an upper limit for the X-ray count rate, we assume a radius $R_1 \sim
4.5 \times 10^{10}$\,cm, which correctly reproduces the amplitude of
the dip in the MACHO light curve. The $3\sigma$ X-ray upper limit of
0.00014 cts/s for the X-ray off state requires a radius of $R_1 \ga 1
\times 10^{10}$\,cm (using $L = 10^{38}$\,erg\,s$^{-1}$, $T \la
185000$\,K, $n_{\rm h} = 6 \times 10^{20}$\,cm$^{-2}$) during the
optical bright state. Our calculations show that it is relatively easy
to reproduce the observed optical dips from the X-ray data, with the
correct amplitude and surprisingly accurate absolute magnitudes. It
also illustrates that when the X-rays become detectable, the white
dwarf photosphere has almost reached its minimum size and the optical
light curve has almost reached the level of the faint phase
plateau. The difference between the observed and predicted optical
light curve immediately after the optical decline could be explained
by the initial lack of an optically thick inner disk after the white
dwarf envelope has contracted to its minimal proportions.

\section{A limit-cycle model for \rxj{}}

The X-ray and optical light curves of \rxj\ suggest that the system follows
a kind of a limit cycle behaviour with four typical time-scales: the
$\sim$\,140 days of the optical high/ X-ray off state,
%
%
the rapid transition ($\la$\,4 days) to the X-ray on/ optical low
state, the $\sim$\,30 days duration of this state,
%
%
and, again, the rapid transition ($\la$\,2 days) to the optical high/
X-ray off state.

We propose here that this behaviour results from expansion of the
white dwarf photosphere in response to enhanced accretion onto the
white dwarf, together with the reaction of the disk to increased
irradiation by this expanded photosphere while the mass transfer from
the companion star remains constant.
%
%
At the (arbitrary) start of the cycle (Fig. \ref{scenario}(d)), let us
assume we have an accretion disk supplying matter to a white dwarf
with its non-expanded radius $R_1 \sim 10^9$~cm. Because the mass
supply rate is close to the Eddington critical accretion rate
$\dot{M}_{\rm crit}$ (Fujimoto 1982; Kato 1985), the white dwarf
radius begins to expand (Fig. \ref{scenario} (d)--(e)), as explained
in Sect. 2 above.
This will, in turn, influence the disk temperature. An extended
central source with radius $R_1 \gg H$, where $H$ is the scale height
of the disk, produces a surface temperature $T_{\rm irr}$ at disk
radius $R$ in an optically thick disk (e.g. Adams et al. 1988) given by
\begin{equation}
\biggl({T_{\rm irr}\over T_1}\biggr)^4
 = {\eta\over\pi}\biggl[\arcsin \rho
 -\rho(1 - \rho^2)^{1/2}\biggr].
\label{eq:tex}
\end{equation}
Here $\rho = R_1/R$, $T_1$ is the temperature of the white dwarf
photosphere, limb-darkening has been neglected, and $\eta$ is the
reprocessing efficiency of the disk surface. 
Increasing $R_1$ has two effects: (i) at given radius $R$, the disk
temperature rises approximately as \mbox{$T_{\rm irr}^4 \propto
R_1\,R^{-3} \propto R_1$}, and (ii) the inner disk disappears in the
hot envelope of the star (see also Sect. 4 above).

The increase in disk temperature raises the mass-flow rate in the
disk, since the disk viscosity coefficient $\nu = \alpha c_{\rm S}H$
is increased. Here $c_{\rm S}$ is the sound speed, and $H = c_{\rm
S}R^{3/2}/(2GM)^{1/2}$ is the scale height of the disk. The disk is
now no longer in a steady state, since the mass-flow rate within it
exceeds the mass supply rate from the companion star at its outer
edge. Its mass is therefore gradually drained onto the white dwarf on
a viscous timescale 
\begin{equation}
t_{\rm visc,d} = {R_d^2\over \nu} = {R_d\over \alpha}{R_d\over {
c_{\rm S}H}} \sim 130\ \left({\alpha\over
0.1}\right)^{-1}\left({R_d\over {10^{11}{\rm
cm}}}\right)~{\rm days},~~~
\label{eq: thot}
\end{equation}
where $10^{11}$\,cm is a characteristic disk radius (the radius of the
Roche lobe is about $1.5\times 10^{11}$\,cm), $H/R_{\rm d} \simeq
0.03$, $c_{\rm S} \simeq 3\times 10^6$\,cm\,s$^{-1}$ for $T_{\rm irr}
\simeq 30\,000$\,K , and $\alpha \simeq 0.1$ is a typical value of the
viscosity parameter. With the disk being drained, the accretion rate
onto the white dwarf eventually drops below $\dot{M}_{\rm crit}$, and
the white dwarf reverts to its unexpanded state, the disk becomes
cooler, and the system enters an optical low and X-ray on state
(Fig. \ref{scenario}\,(a)--(b)). The collapse of the expanded stellar
envelope leaves the accretion disk with an inner hole of approximately
the envelope radius which is gradually refilled by accretion from the
outer disk. The disk temperature at a given radius $R$ decreases as
\mbox{$T_{\rm irr}^4 \propto R_1\,R^{-3}$}, but with $R_1 \sim
10^9$\,cm in the optical low state the temperature at the edge of the
hole ($R_{\rm h}\simeq 3\times 10^{10}$\,cm) becomes $\sim 32\,000$\,K,
similar to the outer disk temperature in the optical high state. The
viscous time scale for refilling the hole, assuming again $H/R_{\rm h}
\simeq 0.03$ and $c_{\rm S} \simeq 3\times 10^6$\,cm\,s$^{-1}$ then is
\begin{equation}
t_{\rm visc,h} = {R_{\rm h}^2\over \nu} = {R_{\rm h}\over
\alpha}{R_{\rm h}\over { c_{\rm S}H}} \sim 40\ \left({\alpha\over
0.1}\right)^{-1}\left({R_{\rm h}\over {3\times 10^{10}{\rm
cm}}}\right){\rm days}.~~~
\label{eq: tcold}
\end{equation}
This picture predicts a long X-ray off state and a
shorter X-ray on state, with rapid (thermal-timescale) transitions
between them, in quantitative agreement with what is observed.

\section{Short-term variability}

\begin{figure}
 \resizebox{\hsize}{!}{\includegraphics{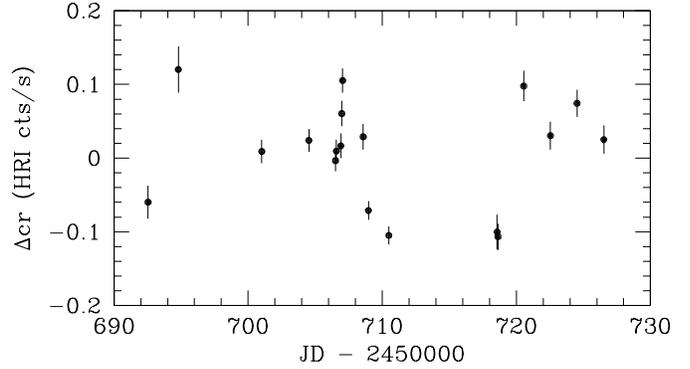}}
\caption[]{\label{hri_residuals}
Observed variability of the soft X-ray flux of \rxj\ during the X-ray 
on-state after subtraction of the exponential long-term flux decline shown 
in Fig. \ref{hri_opt}. }
\end{figure}

Besides the exponential decline, the soft X-ray flux of \rxj\ shows 
significant variations of $\sim \pm$\,0.1 HRI counts/s on timescales of
hours to days (Fig. \ref{hri_residuals}). A time-series analysis of the 
detrended HRI count rates, however, reveals no clear periodicity in the 
range 0.1--10\,days. The strongest signal is found at $P = 0.4075$\,days 
but corresponds only to a $2\sigma$ detection. We have arbitrarily 
phase-folded the residual fluxes on the suggested orbital period 
of $\sim$\,0.76 days but find no obvious modulation of the light curve 
neither using the spectroscopic ephemeris (Crampton et al. 1996) nor using 
the better defined photometric ephemeris (Alcock et al. 1996). 


Although our analysis shows that variability on the orbital and possibly
shorter timescales may be present, our data coverage is not sufficient to
decide whether the flux variations are truly periodic or not. 

\section{Conclusions}

The ROSAT HRI monitoring of a complete X-ray outburst of the transient SSS
\rxj\ has shown that the transition between the X-ray on and off states
occurs with a change of the soft X-ray flux by a factor of $> 100$
within 2--4 days. In the model of an expanding and contracting white dwarf
envelope this implies that the decrease of the effective radius by 
a factor of $>7$ and the increase of the effective temperature by a factor of
$\ga$ 3 occur on the same time-scale during the X-ray turn-on and vice
versa during the X-ray turn-off. 

The steepest intensity variations in the optical occur on a similar
time-scale as the soft X-ray flux variations. This is consistent with our
model that the optical variability is caused by the varying contribution of
the accretion disk illumination by the expanding and contracting envelope
of the white dwarf. 

The existence of typical time-scales of the optical high/ X-ray off state,
of the optical low/ X-ray on state, and of the transition phases, and the
fairly accurate repetition of the optical lightcurve suggests that the observed
variability is driven by limit-cycle behaviour. A possible self-maintained
mechanism is the periodic change of the accretion disk viscosity in
response to changes of the irradiation by the hot central star. In this
scenario, the mass-flow rate at the surface of the white dwarf varies while
the mass-transfer rate from the companion star remains constant. Our model
can qualitatively explain the observed time-scales and requires no external
mechanism like the episodic occurrence of star spots near the $L_1$ point
to trigger the transition between the X-ray on and off states. 

\begin{acknowledgements} 
We thank Will Sutherland (Oxford) for providing some information about the
onset of the August 1997 optical low-state of \rxj . We also thank the
ROSAT team at the MPE (Garching) for their support with the time-critical
scheduling of our HRI observations and for including additional
target-of-opportunity pointings during the final phase of the X-ray
outburst. This work was supported in part by the DLR under grant
50\,OR\,96\,09\,8. 
\end{acknowledgements}

\end{document}